\documentclass[12pt]{article}
\usepackage{latexsym,amsfonts,amsmath,amsthm,amssymb}

\newcommand{\la}{\lambda}

\newcommand{\Om}{\Omega}
\newcommand{\om}{\omega}
\newcommand{\F}{\cal {F}}
\newcommand{\A}{\cal {A}}
\newcommand{{\bP}}{\bf {P}}

\title{Representation of the contextual statistical model by hyperbolic amplitudes}
\author{Andrei Khrennikov\footnote{International Center for Mathematical Modeling
in Physics and Cognitive Sciences, Email: Andrei.Khrennikov@msi.vxu.se; supported by EU-Network
 "QP and Applications} \\
MSI, University of V\"axj\"o, S-35195, Sweden}

\begin{document}

\maketitle

\begin{abstract}
We continue the development of a so called contextual statistical
model (here context has the meaning of a complex of physical
conditions). It is shown that, besides contexts producing the
conventional trigonometric $\cos$-interference, there exist
contexts producing the hyperbolic $\cos$-interference. Starting
with the corresponding interference formula of total probability
we represent such contexts by hyperbolic probabilistic amplitudes
or in the abstract formalism by normalized vectors of a hyperbolic
analogue of the Hilbert space. There is obtained a hyperbolic
Born's rule. Incompatible observables are represented by
noncommutative operators. This paper can be considered as the
first step towards hyperbolic quantum probability. We also discuss
possibilities of experimental verification of hyperbolic quantum
mechanics: in physics of elementary particles, string theory as
well as in experiments with nonphysical systems, e.g. in
psychology, cognitive sciences and economy.
\end{abstract}

\section{Introduction}

In Ref. 1 there was presented a so called contextual viewpoint of the origin of quantum
(conditional) probabilities (here a {\it context} has the meaning of a complex of physical conditions).
Such an approach gives the possibility to unify
classical Kolmogorov (measure theoretical) and quantum (Hilbert space) probability
theories by constructing a natural representation of the Kolmogorov model in a complex Hilbert space.
Thus in the contextual approach quantum probabilistic behavior (in particular, {\it interference of
probabilities}) is simply a consequence of a very special representation of Kolmogorov probabilities --
by complex amplitudes (vectors in a complex Hilbert space). Each representation is based
on a fixed pair of observables (Kolmogorov random variables) $a$ and $b$ -- {\it reference
observables} -- which produce the contextual image of a Kolmogorov probability space in
a complex Hilbert space. The crucial  point is that all Kolmogorov probabilities should be considered
as conditional (or better to say contextual) probabilities, cf. L. Accardi$^{2-4}$, L. Ballentine$^{5,6},$
W. De Muynck$^{7, 8},$ S. Gudder$^{9, 10},$ A. Lande$^{11},$  G. Mackey$^{12}.$

 In Ref.1 we introduced a class ${\cal C}^{\rm{tr}}$ of contexts (``trigonometric contexts'')
 which can be represented by complex probabilistic amplitudes inducing the representation in the complex
 Hilbert space. The ${\cal C}^{\rm{tr}}$ consists of context producing the conventional trigonometric
 $\cos$-interference. However, in general the set of contexts  is not reduced
 to the class of trigonometric contexts ${\cal C}^{\rm{tr}}.$ There exist contexts producing the
 hyperbolic
 $\cosh$-interference. The set of hyperbolic contexts is denoted by the symbol ${\cal C}^{\rm{hyp}}.$

 In this paper we show that it is possible to represent contexts belonging to ${\cal C}^{\rm{hyp}}$
 by so called {\it hyperbolic amplitudes.} Such amplitudes take values in the set of ``hyperbolic numbers''
 (two dimensional Clifford algebra). It will be demonstrated that in the hyperbolic framework
 we can proceed quite far in the same directions as in the trigonometric framework. We obtain hyperbolic analogues
 of the interference of probabilities, probability amplitudes, Born's rule, representation of incompatible observables
 by noncommuting operators... The crucial difference between two representations is that in the hyperbolic case
 the principle of superposition is violated.

\section{Contextual viewpoint to the Kolmogorov model and interference of probabilities}

In this section we repeat the main points of contextual measure-theoretical approach
to interference of probabilities, see Ref. 1 for details.

Let $(\Om, \F, {\bP})$ be a Kolmogorov probability space: $\Om$ is an arbitrary set,
$\F$ is a $\sigma$-field of subsets
of $\Om$ and ${\bP}$ is a countably additive measure on $\F$ taking values in $[0,1]$ and normalized by one
(Kolmogorov probability).
By the standard Kolmogorov axiomatics sets $A\in {\cal F}$ represent {\it{events.}}
In our simplest model of {\it{contextual probability}} (which can be called the
 Kolmogorov contextual space)
the same system of sets, ${\cal F}$, is used to represent complexes of experimental
physical conditions -- {\it{contexts.}}

{\small Thus depending on circumstances a set $O \in {\cal F}$
will be interpreted either as event or as context. We shall sharply distinguish events and
contexts on phemenological level, but we shall use the same
 mathematical object ${\cal F}$ to represent both events and contexts in a mathematical model.
 In principle, in a mathematical model events and contexts can be represented
 by different families of sets, e.g., in Renye's model.
 We will not do this from the beginning. But later we will fix  families
 of contexts, e.g., ${\cal C}^{\rm{tr}}$ or  ${\cal C}^{\rm{hyp}},$ which are
 proper subfamilies of ${\cal F}.$

 The conditional probability is mathematically
 defined by the Bayes' formula: ${\bf P}(A/C)={\bf P}(AC)/{\bf P}(C), {\bf P}(C) \ne 0.$
In our contextual model this probability has the meaning of the probability of occurrence of the event
$A$ under the complex of physical conditions $C.$
Thus it would be more natural to call ${\bf P}(A/C)$ a {\it{contextual probability}}
and not {\it conditional probability.} Roughly speaking to find ${\bf P}(A/C)$ we should
find parameters $\omega^A$ favoring for the occurrence of the event $A$ among parameters
$\omega^C$ describing the complex of physical conditions $C.$

Let ${\A}=\{ A_n \}$ be finite or countable
{\it complete group of disjoint  contexts} (or in the event-terminology
 -- complete group of disjoint events):
$$
A_i A_j= \emptyset, i \not= j,\;\;\;\;  \cup_i A_i=\Om.$$
Let $B\in {\cal F}$  be an event and $C\in {\cal F}$ be a context and let ${\bP}(C)>0.$
We have the standard {\it formula of total probability:}
${\bP}(B/C)=\sum_n {\bP}(A_n/C){\bP}(B/A_nC).$ Let $a=a_1,...,a_n$
and $b=b_1,...,b_n$ be discrete random variables. Then
\begin{equation}
\label{DDK}
{\bP}(b=b_i/C)=\sum_n {\bP}(a=a_n/C) {{\bP}(b=b_i/a=a_n, C)} \;.
\end{equation}
{\small We remark that sets
\begin{equation}
\label{DD}
B_x=\{\om \in \Om: b(\om)=x\}\; \;  \mbox{and} \; \;
A_y=\{\om \in \Om: a(\om)=y\}
\end{equation}
have two different interpretations. On the one hand, these sets represent events corresponding to occurrence
of the values $b=x$ and $a=y,$ respectively. On the other hand, they represent contexts
(complexes of physical conditions) corresponding to selections
of physical systems with respect to values $b=x$ and $a=y,$ respectively. The main problem
with the formula of total probability is that in general it is impossible to construct
a context ``$A_y C$" corresponding to a selection with respect to the value $a=y$ which would not disturb
systems prepared by the context $C.$ But only in the absence of disturbance we can use the set
theoretical operation of intersection. I would like to modify the formula of total
probability by
eliminating sets ``$A_y C$" which in general do not represent physically realizable contexts.}

A set $C$ belonging to $\F$ is said to
be a {\it non degenerate context} with respect to  ${\A}=\{A_n\}$ if
${\bP}(A_nC)\not =0$ for all $n.$ We denote the set of such contexts by the symbol  ${\cal C}_{\A, \rm{nd}}.$

Let ${\A}=\{A_n\}$ and ${\cal B}=\{B_n \}$ be two
complete groups of disjoint  contexts. They are said to be {\it incompatible} if
${\bP}(B_n A_k) \not = 0$ for all $n$ and $k.$
Thus ${\cal B}$ and ${\cal A}$ are incompatible iff every $B_n$ is a non degenerate context with respect to
${\cal A}$ and vice versa. Random variables $a$ and
$b$ inducing, see (\ref{DD}), incompatible complete groups ${\cal A}=\{A_n\}$ and ${\cal
B}=\{B_k\}$ of disjoint  contexts are said to be {\it {incompatible random
variables.}}

{\bf Theorem 2.1.} (Interference formula of total probability)
 {\it Let
 ${\cal A}=\{A_1, A_2= \Omega\setminus A_1\}$ and ${\cal B}=\{B_1,B_2= \Omega\setminus B_1\}$
 be incompatible and let a context $C \in {\cal C}_{\A, \rm{nd}}.$
 Then, for any $B\in {\cal B}:$
%\begin{equation}
%\label{INN}
$$
{\bP}(B/C)=\sum_{j=1}^2 {\bP}(A_j/C){\bP}(B/A_j)+
2 \lambda(B/{\A},C)\sqrt{\prod_{j=1}^2{\bP}(A_j/C)
{\bP}(B/A_j)}
$$
%\end{equation}
where}
$$
\lambda(B/{\cal A}, C)=
\frac{{\bf P}(B/C) - \sum_{j=1}^{2} {\bf P}(B/A_j){\bf P}(A_j/C)}{2\sqrt{{\bP}(A_1/C) {\bP}(B/A_1){\bP}(A_2/C) {\bP}(B/A_2)}}
$$
To prove Theorem, we put the expression for $\lambda$ into
the sum and obtain identity. In fact, this formula is just a representation of
the probability ${\bP}(B/C)$ in a special way.
The $\la(B/{\A},C)$ are called {\it coefficients of
statistical disturbance.} We shall use at few occations teh following result:

{\bf Lemma 2.1.} {\it{Let conditions of Theorem 2.1 hold true. Then}}
\begin{equation}
\label{CD2}
\sum_k \la(B_k/{\A},C)\sqrt{{\bP}(A_1/C){\bP}(A_2/C)
{\bP}(B_k/A_1) {\bP}(B_k/A_2)}=0
\end{equation}

{\bf Proof.} We have
$$
1=\sum_k {\bP}(B_k/C)=\sum_k \sum_n {\bP}(A_n/C)
{\bP}(B_k/A_n)
$$
$$
+\sum_k \la(B_k/{\A},C)\sqrt{{\bP}(A_1/C){\bP}(A_2/C)
{\bP}(B_k/A_1) {\bP}(B_k/A_2)}.
$$
But $\sum_n(\sum_k {\bP}(B_k/A_n)) {\bP}(A_n/C)=1.$

\medskip

1). Suppose that for every $B\in {\cal B}$
$ \vert \la(B/{\A},C)\vert\leq 1 \;.$
In this case we can introduce new statistical parameters $\theta(B/{\A},C)\in
[0,2 \pi]$ and represent the coefficients of statistical disturbance in the
trigonometric form:
$\la(B/{\A},C)=\cos \theta (B/{\A},C).$
Parameters $\theta(B/{\A},C)$ are said to be {\it{relative phases}} of
an event $B$ with respect to  ${\A}$
(in the context $C$). We have the following interference formula of total probability:
%\begin{equation}
%\label{TNC}
$$
{\bP}(B/C)=\sum_{j=1}^2 {\bP}(A_j/C){\bP}(B/A_j)+
2 \cos \theta(B/{\A},C)\sqrt{\prod_{j=1}^2{\bP}(A_j/C){\bP}(B/A_j)}
$$%\end{equation}
This is nothing other than the famous {\it formula of interference of
probabilities.}

In Ref. 1 there was shown that by starting with this formula we can construct the representation of the set
of trigonometric contexts
$$
{\cal C}^{\rm tr}=\{C\in{\cal C}_{\rm{a, nd}}:|\la(B_j/a, c)|\leq 1, j=1,2\}
$$
in the complex Hilbert space, obtain Born's rule and represent incompatible variables $a$ and $b$
by (noncommutative) operators.

2). Suppose that for every $B\in {\cal B},$
$\vert \la(B/{\A},C)\vert\geq  1.$
In this case we can introduce new statistical parameters $\theta(B/{\A},C)\in
(-\infty ,+ \infty)$ and represent the coefficients of statistical disturbance in the
trigonometric form:
$
\la(B/{\A},C)=\pm \cosh \theta(B/{\A},C).
$
Parameters $\theta(B/{\A},C)$ are said to be hyperbolic {\it{relative phases.}}
In this case we obtain the formula of total probability with hyperbolic $\cosh$-interference:
\begin{equation}
\label{INTHYP}
{\bP}(B/C)=\sum_{j=1}^2 {\bP}(A_j/C){\bP}(B/A_j)\pm
2 \cosh \theta(B/{\A},C)\sqrt{\prod_{j=1}^2{\bP}(A_j/C){\bP}(B/A_j)}
\end{equation}
The aim of this paper is to show that by starting with this formula we can construct the representation of the set
of hyperbolic contexts
$$
{\cal C}^{\rm hyp}=\{C\in{\cal C}_{\rm{a, nd}}:|\la(B_j/a, c)|\geq 1, j=1,2\}
$$
in the hyperbolic Hilbert space, obtain an analogue of Born's rule and represent incompatible variables $a$ and $b$
by (noncommutative) operators.

We can also consider the case of mixed hyper-trigonometric behavior: one of coefficients
is larger than 1 and one is smaller than 1. However, in this paper we shall discuss only the
case of the hyperbolic interference.

In our further considerations the complete groups of  disjoint  contexts
${\cal A}$ and ${\cal B}$ will correspond to some incompatible random variables
$a$ and $b.$ We shall use the symbols  $\lambda(b=x/a, C)$
instead of $ \lambda(b=x/{\cal A}, C).$

\section{Representation of contexts by hyperbolic amplitudes, hyperbolic Hilbert space representation.}

Everywhere below we study  contexts producing the {\bf hyperbolic interference} for incompatible dichotomous random
variables $a=a_1, a_2, b=b_1, b_2.$ This pair of variables will be fixed.
We call such variables {\bf reference variables.} For each pair $a, b$ of reference variables
we construct a representation of the set of contexts ${\cal C}^{\rm hyp}$ in
hyperbolic Hilbert space (``quantum-like representation'').

{\bf 3.1. Hyperbolic algebra.}
Instead of the field complex numbers
${\bf C},$  we shall use  so called {\bf hyperbolic numbers,} namely the two dimensional
Clifford algebra, ${\bf G}.$ We call this algebra {\it hyperbolic algebra.}

{\bf Remark 3.1.} {\small Of course, it is rather dangerous to invent an own name for a notion established almost as
firm as complex numbers. We use a new name,  hyperbolic algebra, for the well known algebraic
object, the two dimensional Clifford algebra, by following reasons. First we explain why we dislike
to use the standard notion Clifford algebra in this particular case. The standard Clifford machinery
was developed around noncommutative features of general Clifford algebras. The two dimensional Clifford
algebra, hyperbolic algebra in our terminology, is commutative. Commutativity of ${\bf G}$ is very important
in our considerations. We now explain why we propose the name hyperbolic algebra.
Hyperbolic functions are naturally related to the algebraic structure
of ${\bf G}$ through a hyperbolic generalization of Euler's formula for the complex
numbers. This is the crucial point of our considerations - the possibility to use this
algebraic structure to represent some special transformations for hyperbolic functions.}

Denote by the symbol $j$  the generator of the algebra ${\bf G}$ of hyperbolic numbers:
$$
j^2=1.
$$
The algebra {\bf{G}} is the two dimensional real
algebra with basis $e_0=1$ and $e_1=j.$  Elements of {\bf{G}} have the form $z=x + j y, \; x, y \in {\bf{R}}.$
We have $z_1 + z_2=(x_1+x_2)+j(y_1+y_2)$ and $z_1 z_2=(x_1x_2+y_1y_2)+j(x_1y_2+x_2y_1).$
This algebra is commutative. It is not a field - not every element has the inverse one.

We introduce an involution in {\bf{G}} by setting
$\bar{z} = x - j y$
and set  $|z|^2=z\bar{z}=x^2-y^2.$
We remark that  $|z|=\sqrt{x^2-y^2}$ {\bf is not well defined} for an arbitrary $z\in {{\bf{G}}}.$
We set ${{\bf{G}}}_+=
\{z\in{{\bf{G}}}:|z|^2\geq 0\}.$ We remark that ${{\bf{G}}}_+$
is a multiplicative semigroup as follows from the equality

$|z_1 z_2|^2=|z_1|^2 |z_2|^2.$

Thus, for $z_1, z_2 \in {{\bf{G}}}_+,$
we have that $|z_1 z_2|$ is well defined and
$|z_1 z_2|=|z_1||z_2|.$ We define a hyperbolic exponential function by using
a hyperbolic analogue of the Euler's formula:

$e^{j\theta}=\cosh\theta+ j \sinh\theta, \; \theta \in {\bf{R}}.$

We remark that

$e^{j\theta_1} e^{j\theta_2}=e^{j(\theta_1+\theta_2)}, \overline{e^{j\theta}}
=e^{-j\theta}, |e^{j\theta}|^2= \cosh^2\theta - \sinh^2\theta=1.$

Hence, $z=\pm e^{j\theta}$ always belongs to ${{\bf{G}}}_+.$
We also have

$\cosh\theta=\frac{e^{j\theta}+e^{-j\theta}}{2}, \;\;\sinh\theta=\frac{e^{j\theta}-e^{-j\theta}}{2 j}\;.$

We set ${{\bf{G}}}_+^*=
\{z\in{{\bf{G}}}_+:|z|^2>0 \}. $
Let  $z\in {{\bf{G}}}_+^*.$  We have

$z=|z|(\frac{x}{|z|}+j \frac{y}{|z|})= \rm{sign}\; x\; |z|\;(\frac{x {\rm{sign}} x}{|z|} +j\;
\frac{y {\rm{sign}} x}{|z|}).$

As $\frac{x^2}{|z|^2}-\frac{y^2}{|z|^2}=1,$  we can represent $x$ sign $x= \cosh\theta$
and $y$ sign $x=\sinh\theta, $ where the phase $\theta$ is unequally defined.
We can represent each $z\in {{\bf{G}}}_+^*$ as

$z = \rm{sign}\; x\;  |z|\; e^{j\theta}\;.$

By using this representation we can easily prove that ${{\bf{G}}}_+^*$
is a multiplicative group. Here $\frac{1}{z}=\frac{{\rm{sign}} x}{|z|}e^{-j\theta}.$
The unit circle in ${{\bf{G}}}$ is defined as $S_1 = \{z\in{{\bf{G}}}:|z|^2=1\}
=\{ z= \pm e^{j \theta}, \theta \in (-\infty, +\infty)\}.$ It is a multiplicative
subgroup of ${\bf G}_+^*.$

To construct a ${\bf{G}}$-linear representation of the set ${\cal C}^{\rm hyp}$
of hyperbolic contexts, we shall use the following elementary formula:
\begin{equation}
\label{EEE}
D=A+B\pm 2AB\cosh \theta=|\sqrt{A}\pm e^{j\theta}\sqrt{B}|^2,
\end{equation}
for real coefficients $A, B>0.$

{\bf 3.2. Hyperbolic probability amplitude, hyperbolic Born's rule.}
We set $Y=\{a_1, a_2\}, X=\{b_1,
b_2\}$ (``spectra'' of random variables $a$ and $b).$
Let $C\in {\cal C}^{\rm hyp}.$ We set
\[p_C^a(y)={\bP}(a=y/C), p_C^b(x)={\bP}(b=x/C), p(x/y)={\bP}(b=x/a=y),\]
$x \in X, y \in Y.$
The interference formula of total probability (\ref{INTHYP}) can be written in the following form:
\begin{equation}
\label{TwoH}p_C^b(x)=\sum_{y \in Y}p_C^a(y) p(x/y) \pm 2\cosh \theta_C(x)\sqrt{\Pi_{y \in
Y}p_C^a(y) p(x/y)}\;,
\end{equation}
where $\theta_C(x)=\theta(b= x/a, C) = \pm \rm{arccosh} \vert \lambda(b=x/a, C)\vert,
x \in X, C \in {\cal C}^{\rm hyp}.$ Here the coefficient $\lambda$ is defined by
\begin{equation}
\label{Two1}\lambda(b=x/a, C)  =\frac{p_C^b(x) - \sum_{y \in Y}p_C^a(y) p(x/y)}{2\sqrt{\Pi_{y\in Y}p_C^a(y)p(x/y)}}.
\end{equation}
 By using (\ref{EEE}) we can represent the probability $p_C^b(x)$ as the square of the
hyperbolic amplitude:
\begin{equation}
\label{BORNH}
p_C^b(x)=|\varphi_C(x)|^2,
\end{equation}
where
\begin{equation}
\label{TwoA}
\varphi(x) \equiv \varphi_C(x)= \sqrt{p_C^a(a_1)p(x/a_1)} + \epsilon_C(x)
e^{j\theta_C(x)} \sqrt{p_C^a(a_2)p(x/a_2)} \;.
\end{equation}
Here $\epsilon_C(x)={\rm sign}\; \lambda(x/a, C).$ We remark that by Lemma 2.1:
\begin{equation}
\label{MART}
\sum_{x\in X}\epsilon_C(x)=0.
\end{equation}
Thus we have a {\it hyperbolic generalization of Born's rule} for the $b$-variable, see (\ref{BORNH}).

{\bf 3.3. Hyperbolic Hilbert space.} Hyperbolic Hilbert space is
${{\bf{G}}}$-linear space (module) ${\bf{E}}$
with a ${{\bf{G}}}$-linear scalar
product: a map $(\cdot, \cdot): {\bf{E}}\times {\bf{E}} \to {{\bf{G}}}$ that is

1) linear with respect to the first argument:

$ (a z+ b w, u) = a (z,u) + b (w, u), a,b \in {{\bf{G}}},
z,w, u \in {\bf{E}};$

2) symmetric: $(z,u)= \overline{(u,z)} ;$

3) nondegenerate: $(z,u)=0$ for all $u \in {\bf{E}}$ iff $z=0.$

{\bf Remark 3.2.} If we consider ${\bf{E}}$ as just a ${\bf R}$-linear space, then $(\cdot, \cdot)$
is a bilinear form which is not positive defined.
In particular, in the two dimensional case we have the signature: $(+,-,+,-).$

{\bf 3.4. Hyperbolic Hilbert space representation.} We introduce on the space
$\Phi(X, {\bf G})$ of functions: $\varphi: X\to {\bf G}.$
Since $X= \{b_1, b_2 \},$ the $\Phi(X, {\bf G})$ is the two dimensional ${\bf G}$-module.
We define the ${\bf G}$-scalar product by
\begin{equation}
\label{BHS}
(\varphi, \psi) = \sum_{x\in X} \varphi(x)\bar \psi(x).
\end{equation}
with conjugation in the algebra ${\bf G}.$
The system of functions $\{e_x^b\}_{x\in X}$ is an orthonormal basis in the hyperbolic
Hilbert space $H^{\rm hyp}=(\Phi(X, {\bf G}), (\cdot, \cdot)).$
Thus we have the hyperbolic analogue of the Born's rule in $H^{\rm hyp}:$
\begin{equation}
\label{BH}
p_C^b(x)=\vert(\varphi_C, e_x^b)\vert^2 \;.
\end{equation}

Let $X \subset R.$ By using the hyperbolic Hilbert space representation ({\ref{BH}}) of
the Born's rule  we obtain  the hyperbolic Hilbert space representation of the
expectation of the (Kolmogorovian) random variable $b$:
\begin{equation}
\label{BI1}
E (b/C)= \sum_{x\in X}xp_C^b(x)=\sum_{x\in X}x\vert\varphi_C(x)\vert^2=
\sum_{x\in X}x (\varphi_C, e_x^b) \overline{(\varphi_C, e_x^b)}=
(\hat b \varphi_C, \varphi_C) \;,
\end{equation}
where  the  (self-adjoint) operator $\hat b: H^{\rm hyp} \to H^{\rm hyp}$ is determined by its
eigenvectors: $\hat b e_x^b=x e^b_x, x\in X.$
This is the  multiplication operator in the space of ${\bf G}$-valued functions $\Phi(X,{\bf {\bf G}}):$
$$
\hat{b} \varphi(x) = x \varphi(x)
$$
 By (\ref{BI1}) the  conditional expectation of the Kolmogorovian
random variable $b$ is represented
with the aid of the self-adjoint operator $\hat b.$

Thus we constructed a ${\bf G}$-linear representation of the contextual Kolmogorov model:
$$
J^{b/a}: {\cal C}^{\rm{hyp}} \to H^{\rm hyp}.
$$
We set $S_{{\cal C}^{\rm{hyp}}} = J^{b/a} ({\cal C}^{\rm{hyp}} ).$ This is a subset of the unit sphere
$S$ of the Hilbert space $H^{\rm hyp}.$ We introduce the coefficients
\begin{equation}
\label{KOE}
u_j^a=\sqrt{p_C^a(a_j)}, u_j^b=\sqrt{p_C^b(b_j)}, p_{ij}=p(b_j/a_i), u_{ij}=\sqrt{p_{ij}},
\theta_j=\theta_C(b_j).
\end{equation}
and $\epsilon_i=\epsilon(b_i).$
We remark that the coefficients $u_j^a, u_j^b$ depend on a context $C;$ so
$u_j^a=u_j^a(C), u_j^b=u_j^b(C).$
We also consider the {\it matrix of transition
probabilities} ${\bf P}^{b/a}=(p_{ij}).$ It is always a {\it stochastic matrix:}
$p_{i1}+ p_{i2}= 1, i=1,2.$ In further considerations we shall also consider
{\it double stochastic} matrices: $p_{1j}+ p_{2j}= 1, j=1,2.$

We represent a state $\varphi_C$ by $\varphi_C=v_1^b e_1^b + v_2^b e_2^b,$
where $v_i^b=u_1^a u_{1i} + \epsilon_i u_2^a u_{2i} e^{j\theta_i}.$
So
$$
p_C^b(b_i)=|v_i^b|^2=|u_1^a u_{1i} + \epsilon_i u_2^a u_{2i} e^{j\theta_i}|^2 \;.
$$
This is the {\bf G-linear representation of the hyperbolic
interference of probabilities.} This formula can also be derived
in the formalism of the hyperbolic Hilbert space, see section 4.
We remark that here the ${\bf G}$-linear combination $u_1^a u_{1i}
+ \epsilon_i u_2^a u_{2i} e^{j\theta_i}$ belongs to
${{\bf{G}}}_+^*.$

Thus for any context $C_0\in{\cal C}^{\rm hyp}$ we can represent $\varphi_{C_0}$ in the form:
\[\varphi_{C_0}=u_1^a e_1^a + u_2^a e_2^a,\] where
$$
e_1^a=(u_{11}, u_{12}) \; ,
e_2^a=(\epsilon_1 e^{j\theta_1} u_{21}, \epsilon_2 e^{j\theta_2} u_{22}).
$$
As in the $\bf C$-case$^{1}$, we introduce the matrix $V$ with coefficients
$v_{11}=u_{11}, v_{21}=u_{21}$ and $v_{12}=\epsilon_1e^{j\theta_1} u_{21},
v_{22}=\epsilon_2e^{j\theta_2} u_{22}.$ We remark that here coefficients $v_{ij} \in {{\bf{G}}}_+^*.$
In the same way as in the complex case the Born's rule
\begin{equation}
\label{ME}
p_{C_0}^a (a_i)=|(\varphi_{C_0}, e_i^a)|^2
\end{equation}
holds true in the $a$-basis iff $\{e_i^a\}$ is an orthonormal basis in $H^{\rm hyp}.$
The latter is equivalent to the $\bf G$-unitary of the matrix $V$
(corresponding to the transition from $\{e_i^b\}$ to $\{e_i^a\}): \overline{V}^*V=I,$ or
\begin{equation}
\label{NORM1}
\bar v_{11} v_{11} + \bar v_{21} v_{21}=1,\;
\bar v_{12} v_{12} + \bar v_{22} v_{22}=1,
\end{equation}
\begin{equation}
\label{NORM2}
\bar v_{11} v_{12} + \bar v_{21} v_{22}=0.
\end{equation}
Thus $1=u_{11}^2 + u_{21}^2=p(b_1/a_1) + p(b_1/a_2)$ and $1=u_{12}^2 + u_{22}^2=p(b_2/a_1) + p(b_2/a_2).$
Thus the first two equations of the $\bf G$-unitary are equivalent to the double stochasticity
of ${\bf P}^{b/a}$ (as in the ${\bf C}$-case$^{1}$). We remark that the equations (\ref{NORM1}) can be
written as
\begin{equation}
\label{NORM3}
\vert v_{11}\vert^2 + \vert v_{21}\vert^2 =1, \vert v_{12}\vert^2 + \vert v_{22}\vert^2=1,
\end{equation}
cf. section 4. The third unitarity equation (\ref{NORM2}) can be
written as
\begin{equation}
\label{NORM4}
u_{11} u_{12} \epsilon_1 e^{-j\theta_2} + u_{21} \epsilon_2 e^{-j\theta_2} u_{22}=0.
\end{equation}
By using double
stochasticity of ${\bf P}^{a/b} $ we obtain
$e^{j\theta_1}=e^{j\theta_2}.$ Thus
\begin{equation}
\label{MARE}
\theta_1=\theta_2.
\end{equation}

{\bf Lemma 3.1.}
{\it{Let $a$ and $b$ be incompatible random variables and let ${\bf P}^{b/a}$ be double stochastic. Then
\begin{equation}
\label{Ka}
\cosh \theta_C(b_2)=\cosh \theta_C(b_1)
\end{equation}
for any context $C\in{\cal C}^{\rm hyp}$.}}

{\bf Proof.} By Lemma 2.1 we have:
\[\sum_x\epsilon(x)\cosh \theta_C(x)\sqrt{\Pi_y p_C^a(y) p(x/y)} =0.\]
Double stochasticity of ${\bf P}^{b/a}$ implies (\ref{Ka}).

\medskip

The constraint (\ref{Ka}) induced by double stochasticity
can be written as the constraint to phases:
\begin{equation}
\label{MK}
\theta_C(b_2)=\pm \theta_C(b_1).
\end{equation}
To obtain unitary of the matrix $V$ of transition $\{e_i^b\}\to \{e_i^a\}$ we should choose
phases according to (\ref{MARE}). And by (\ref{MK}) we can always do this for a
double stochastic matrix of transition
probabilities.

By choosing such a representation we obtain the
hyperbolic generalization of the Born's rule for the $a$-variable:
\begin{equation}
\label{BBR}
p_C^a(a_j)=\vert(\varphi, e_j^a)\vert^2 \; .
\end{equation}
We now investigate the possibility to use one fixed
basis $\{e_j^a \equiv e_j^a(C_0)\}, C_0 \in {\cal C}^{\rm hyp},$ for all
states $\varphi_C, C\in {\cal C}^{\rm hyp}.$ For any $C\in {\cal C}^{\rm hyp}$ we
would like to have the representation:
\begin{equation}
\label{LUU}
\phi_C= v_1^a(C) e_1^a(C_0) + v_2^a(C) e_2^a(C_0),\;\; \mbox{where}\;\;\; \vert v_j^a(C) \vert^2= p_C^a(a_j).
\end{equation}

We have
$$\varphi_C(b_1)=u_1^a(C) v_{11}(C_0) +
\epsilon_C(b_1) \epsilon_{C_0}(b_1) e^{j[\theta_C(b_1) - \theta_{C_0}(b_1)]} u_2^a(C) v_{12}(C_0)$$
$$\varphi_C(b_2)=u_1^a(C) v_{21}(C_0) +
\epsilon_C(b_2) \epsilon_{C_0}(b_2) e^{j[\theta_C(b_2) - \theta_{C_0}(b_2)]} u_2^a(C) v_{22}(C_0)$$

Thus to obtain (\ref{LUU}) we should have
\[\epsilon_C(b_1) \epsilon_{C_0}(b_1) e^{j [ \theta_C(b_1) - \theta_{C_0}(b_1)]}=
\epsilon_C(b_2) \epsilon_{C_0}(b_2) e^{j [\theta_C(b_2) - \theta_{C_0}(b_2)]}\]
Thus
\[\;\; \theta_C(b_1)-\theta_{C_0}(b_1)=\theta_C(b_2)-\theta_{C_0}(b_2), \; \rm{or}
\;\; \theta_C(b_1)-\theta_{C}(b_2)=\theta_{C_0}(b_1)-\theta_{C_0}(b_2).\]
By choosing the representation
with (\ref{MARE}) we satisfy the above condition.

\medskip

{\bf Theorem 3.1} {\it We can construct the quantum-like  (Hilbert space) representation of
a contextual Kolmogorov space such that the hyperbolic Born's rule holds true for both reference
variables $a$ and $b$ iff the matrix of transition probabilities ${\bf P}^{b/a}$ is double stochastic.}

We remark that basic
contexts $B_x= \{ \omega\in \Omega: b(\omega) = x\}, x \in  X,$ always belong to ${\cal C}^{\rm hyp},$ so $\varphi_{B_x}\in H^{\rm hyp};$ and
$B_x\in {\cal C}^{\rm tr} \cap {\cal C}^{\rm hyp}$ iff $a$ and $b$ are uniformly distributed
(${\bf P}^{a/b}$ and ${\bf P}^{b/a}$ are double stochastic).

\section{Hyperbolic quantum mechanics}

As in the ordinary quantum formalism,
we represent physical states by normalized vectors of a hyperbolic Hilbert space ${\bf{E}}:$
$\varphi\in {\bf{E}}$ and $(\varphi, \varphi)=1.$
We shall consider only dichotomous physical variables and quantum states belonging to
the two dimensional Hilbert space. Thus everywhere below ${\bf{E}}$ denotes the two dimensional
space. Let $a=a_1, a_2$ and $b=b_1, b_2$ be two
physical variables. We represent they by  ${{\bf{G}}}$-linear operators:
$\hat{a}= \vert a_1> < a_1\vert + \vert a_2> < a_2\vert$ and $\hat{b}=
\vert b_1> < b_1\vert + \vert b_2> < b_2\vert,$
where $\{\vert a_i>\}_{i=1,2}$ and  $\{\vert b_i>\}_{i=1,2}$ are two orthonormal bases
in ${\bf{E}}.$ The latter condition plays the fundamental role in hyperbolic quantum mechanics.
This is an analogue of the representation of physical observables by self-adjoint operators
in the conventional quantum mechanics (in the complex Hilbert space).

Let $\varphi$ be a state (normalized vector belonging to ${\bf{E}}).$ We can perform
the following operation (which is well defined from the mathematical point of view).
We expend the vector $\varphi$ with respect
to the basis
$\{\vert b_i>\}_{i=1,2}:$
\begin{equation}
\label{E1}
\varphi = v_1^b \vert b_1>+ v_2^b \vert b_2>,
\end{equation}
where the coefficients (coordinates) $v_i^b$ belong to ${\bf G}.$
We remark that we consider the two dimensional ${\bf G}$-Hilbert
space. There exists (by definition) a basis consisting of two vectors.
As the basis $\{\vert b_i>\}_{i=1,2}$ is orthonormal, we have (as in the complex case) that:
\begin{equation}
\label{E2}
|v_1^b|^2 + |v_2^b|^2 = 1\;.
\end{equation}
However, we could not automatically use Born's probabilistic interpretation for
normalized vectors in the hyperbolic Hilbert space:
it may be that $v_i^b \not\in {\bf G}_+$ and hence $\vert v_i^b \vert^2 < 0$
(in fact, in the complex case we have
${\bf C}={\bf C}_+$; thus there is no problem with positivity).
Since we do not want to consider negative probabilities, in such a case we cannot use
the hyperbolic version of Born's probability interpretation.

{\bf Definition 4.1.} {\it A state $\varphi$ is {\it decomposable}
with respect  to the system of states  $\{\vert b_i> \}_{i=1,2}$ ($b$-decomposable) if}
\begin{equation}
\label{E3}
v_i^b \in {\bf G}_+ \; .
\end{equation}

In such a case we can use generalization of Born's probabilistic interpretation for
a hyperbolic Hilbert space. Numbers
$$
p_\varphi^b(b_i)= \vert v_i^b \vert^2, i=1,2,
$$
are interpreted as probabilities
for values $b=b_i$ for the ${\bf G}$-quantum state $\varphi.$

{\small We remark that in this framework (here we started with a hyperbolic Hilbert space and not with
a contextual statistical model, cf. section 3) a hyperbolic generalization of Born's rule
is  a postulate!}

Thus decomposability is not a mathematical notion. This is not just linear algebraic
decomposition of a vector with respect a basis. This is a physical notion describing
the possibility of probability interpretation of a measurement over a state. As it was already
mentioned, in hyperbolic quantum mechanics a state $\varphi\in {\bf E}$ is not always decomposable. Thus for an
observable $b$ there can exist a state $\varphi$ such that the probabilities $p_\varphi^b(b_i)$ are not well defined.
One of reasons for this can be the impossibility to perform the $b$-measurement for systems in the state $\varphi.$
Such a situation is quite natural from the experimental viewpoint. Moreover, it looks surprising that in ordinary
quantum (as well as classical) theory we can measure any observable in any state. I think that this is just
a consequence of the fact that there was fixed the set of states corresponding to a rather special class of
physical observables. Thus in the hyperbolic quantum formalism fro each state $\varphi \in {\bf E}$
there exists its own set of observables ${\cal O}(\varphi).$ And in general ${\cal O}(\varphi) \not= {\cal O}(\psi).$
We cannot exclude another possibility. The set of observables ${\cal O}$ does not depend on a state $\varphi.$
And the result of an individual measurement of any $b\in {\cal O}$ is well defined for any state $\varphi.$ But
relative frequencies of realizations of the value $b=b_k$ do not converge to any limit. Therefore probabilities
are not well defined. Thus the principle of the statistical stabilization should be violated,
cf. Ref 13.

{\small Let ${\cal K}$ be a Kolmogorov probability model and let
$\varphi \in S_{{\cal C}^{\rm{hyp}}}.$ Thus $\varphi= \varphi_C$ for some context
$C \in {\cal C}^{\rm{hyp}}.$ Let the matrix of transition probabilities ${\bf P}^{b/a}$ be double
stochastic. Then $\varphi$ is decomposable with respect to both reference variables $b$ and
$a.$ Moreover, basis vectors $e_i^b= \vert b_i>$ are $a$-decomposable and vice versa.}

We now start the derivation of the hyperbolic probabilistic rule by using the hyperbolic
Hilbert space formalism. Suppose that a state $\varphi \in {\bf E}$ is $a$-decomposable:
$$
\varphi= v_1^a \vert a_1> + v_2^a \vert a_2>
$$
and the coefficients $v_i^a \in {\bf G}_+.$

We  also suppose that each state $\vert a_i>$ is decomposable with respect
to the system of states $\{\vert b_i>\}_{i=1,2}.$ We have:
\begin{equation}
\label{E4}
\vert a_1>=v_{11} \vert b_1> + v_{12} \vert b_2>,\; \;
\vert a_2>= v_{21} \vert b_1> + v_{22} \vert b_2>\;,
\end{equation}
where the coefficients $v_{ik}$ belong to ${\bf G}_+.$   We have (since both bases are orthonormal):
\begin{equation}
\label{E5}
|v_{11}|^2 + |v_{12}|^2 = 1, \; \;|v_{21}|^2 + |v_{22}|^2 = 1\;,
\end{equation}
cf. (\ref{NORM3}).
We can use the probabilistic interpretation of numbers $ p_{ik} = |v_{ik}|^2,$ namely
$p_{ik}= p_{\vert a_i>}(b_k)$ is the probability for $b=b_k$ in the state $\vert a_i>.$

Let us consider matrix $V=(v_{ik}).$
As in the complex case, the matrix $V$ is unitary, since vectors $\vert a_1>= (v_{11}, v_{12})$
and $\vert a_2>= (v_{21}, v_{22})$ are orthonormal. Hence we have normalization conditions
(\ref{E5}) and the orthogonality condition:
\begin{equation}
\label{EA}
v_{11} \bar{v}_{21} + v_{12} \bar{v}_{22}=0 \;,
\end{equation}
cf. (\ref{NORM2}).
It must be noticed that in general unitarity does not imply that $v_{ik} \in {\bf G}_+.$
The latter condition is the additional constraint on the unitary matrix $V.$
Let us consider the matrix  ${\bf P} ^{b/a} =(p_{ik}).$ This matrix is double
stochastic (since $V$ is unitary).

By using the ${\bf G}$-linear space calculation (the change of the basis) we get
$\varphi= v_1^b \vert b_1> + v_2^b \vert b_2>,$
where $v_1^b = v_1^a v_{11}+ v_2^a v_{21}$ and
$v_2^b = v_1^a v_{12}+ v_2^a v_{22}.$

We remark that decomposability is not transitive. In principle $\varphi$
may be not decomposable with respect to $\{\vert b_i>\}_{i=1,2},$
despite the decomposability of $\varphi$ with respect to $\{\vert a_i>\}_{i=1,2}$
and the decomposability of the latter system with
respect to $\{\vert b_i>\}_{i=1,2}.$

The possibility of decomposability is based on two (totally different) conditions:
(\ref{E2}), normalization, and (\ref{E3}), positivity. Any ${\bf G}$-unitary transformation
preserves the normalization condition. Thus we get automatically that
$\vert v_1^b \vert^2 + \vert v_2^b \vert^2 =1.$ However, the condition of positivity in general is not preserved:
it can be that $v_i^b \not\in {\bf G}_+$ even if we have $v_i^a \in {\bf G}_+$ and
the matrix $V$ is ${\bf G}$-unitary.

Finally,  suppose that $\varphi$ is decomposable with respect to $\{\vert b_i>\}_{i=1,2}.$
Thus  $v_k^b \in {\bf G}_+.$
Therefore coefficients $p_\varphi^b(b_i) = \vert v_i^b \vert^2$ can be interpreted as
probabilities for $b=b_k$ for the ${\bf G}$-quantum state $\varphi.$

Let us consider states such that coefficients $v_i^a, v_{ik}$ belong to ${\bf G}_+^*.$
We can uniquely represent them
as

$v_i^a=\pm \sqrt{p_\varphi^a(a_i)} e^{j \xi_i},
v_{ik}=\pm \sqrt{p_{ik}} e^{j \gamma_{ik}}, i, k, =1,2.$

We find that
\begin{equation}
\label{E7a}
p_\varphi^b(b_1) = p_\varphi^a(a_1) p_{11} + p_\varphi^a(a_2) p_{21} +
2 \epsilon_1 \cosh \theta_1 \sqrt{p_\varphi^a(a_1) p_{11}p_\varphi^a(a_2) p_{21}} \;,
\end{equation}
\begin{equation}
\label{E7b}
p_\varphi^b(b_2) = p_\varphi^a(a_1) p_{12} + p_\varphi^a(a_2) p_{22} +
2 \epsilon_2 \cosh \theta_2 \sqrt{p_\varphi^a(a_1) p_{12} p_\varphi^a(a_2) p_{22}} \;,
\end{equation}
where $\theta_i = \eta+ \gamma_i$  and $\eta= \xi_1- \xi_2,
\gamma_1= \gamma_{11}- \gamma_{21}, \gamma_1= \gamma_{12}- \gamma_{22}$
and $\epsilon_i= \pm.$
To find the right relation between signs of the last terms in equations  (\ref{E7a}),
(\ref{E7b}), we use the normalization condition
\begin{equation}
\label{E7}
\vert v_2^b \vert^2 + \vert v_2^b \vert^2=1
\end{equation}
(which is a consequence of the normalization of $\varphi$ and orthonormality of
the system $\{\vert b_i>\}_{i=1,2}).$

{\small We remark that the normalization condition (\ref{E7}) can be reduced to
relations between coefficients of the transition matrix $V.$ So it does not depend
on the original $a$-decomposition of $\varphi,$ namely coefficients $v_i^a.$ Condition
of positivity, $\vert v_i^b \vert^2 \geq 0,$  could not be written
by using only coefficients of $V.$ We also need to use coefficients $v_i^a.$
Therefore it seems to be impossible to find such a class of linear transformations
$V$ that would preserve condition of positivity, ``decomposition-group" of operators.}

Equation (\ref{E7}) is equivalent to the equation:
\begin{equation}
\label{E8}
\sqrt{p_{12}p_{22}} \cosh\theta_2 \pm \sqrt{p_{11}p_{21}} \cosh\theta_2=0.
\end{equation}
Thus we have to choose opposite signs in equations (\ref{E7a}),
(\ref{E7b}). Unitarity of $V$ also implies that $\theta_1 -
\theta_2 =0,$ so $\gamma_1= \gamma_2.$ We recall that in the
ordinary quantum mechanics we have similar conditions, but
trigonometric functions are used instead of hyperbolic and phases
$\gamma_1$ and $\gamma_2$ are such that $\gamma_1 - \gamma_2=
\pi.$

Finally, we get that unitary linear transformations in the ${\bf G}$-Hilbert space
(in the domain of  decomposable states) represent the following transformation
of probabilities:
\begin{equation}
\label{E7c} p_\varphi^b(b_1) = p_\varphi^a(a_1) p_{11} +
p_\varphi^a(a_2) p_{21} \pm 2 \epsilon_1 \cosh \theta_1
\sqrt{p_\varphi^a(a_1) p_{11}p_\varphi^a(a_2) p_{21}} \;,
\end{equation}
\begin{equation}
\label{E7d} p_\varphi^b(b_2)= p_\varphi^a(a_1) p_{12} +
p_\varphi^a(a_2) p_{22} \mp 2 \epsilon_2 \cosh \theta_2
\sqrt{p_\varphi^a(a_1) p_{12} p_\varphi^a(a_2) p_{22}}  .
\end{equation} This is hyperbolic interference. In section 2 it was
derived  from the contextual statistical model and then in section
3 by using interference formulas we obtained the hyperbolic
Hilbert space representation for contexts. In this section we
started directly from the hyperbolic Hilbert space representation
and derived interference of probabilities.

\section{Experimental verification of hyperbolic quantum mechanics}

This paper contains an important experimental prediction:

\medskip

{\it In statistical experiments with physical (micro as well as
macro) systems there could be produced not only the ordinary
trigonometric, but also the hyperbolic interference picture.}

\medskip

We start with the general description of
interference experiments for discrete observables. There are considered two dichotomous observables:
$a$ - ``slit number'', and $b$ - ``position of a particle on the registration screen.'' The observable
$a$ is measured in the following way. There are placed particle detectors behind the screen having two open slits.
The observable $a=j$ if the detector behind the $j$th slit clicks.  To define
another observable, we choose some domain $D$ on the registration screen and we set $b=1$ if a particle is registered
inside $D$ and $b=0$ if outside. The complex of physical conditions under consideration (context) $C$
is screen with two open slits  and the registration screen. We find frequency
probabilities $p_C^b(1)$ and $p_C^b(0)$ by counting the numbers of particles inside and outside the domain $D$
on the registration screen. Then we perform the measurement of the $a$-variable by placing detectors behind
the first screen. We find frequency probabilities $p_C^a(1)$ and $p_C^a(2)$ by counting the numbers of particles
passing through the first screen and the second screen, respectively (if the source is located symmetrically with respect to
screens, then $p_C^a(a=1) = p_C^a(a=2)=1/2).$ We also find transition probability $p^{b/a}(i/j)$ by closing the $j$th slit  and performing the $b$-measurement under this complex of physical conditions.
For systems described by classical (noncontextual) probability theory
we  get the well known formula of total probability:
$$
p_C^b(x)= p^{b/a}(x/1)p_C^a(1) + p_C^a(2) p^{b/a}(x/2).
$$
Here the coefficient of statistical disturbance $\la(b=x/a,C)=0$. For systems described by quantum  probability,
we get the interference formula:
$$
p_C^b(x)= p^{b/a}(x/1)p_C^a(1) + p_C^a(2) p^{b/a}(x/2) + 2 \cos \theta \sqrt{p^{b/a}(x/1)p_C^a(1) p_C^a(2) p^{b/a}(x/2)}
$$
This formula is usually derived in the Hilbert space formalism. In the book of Feynman an Hibs$^{(14)}$
violation of the formula of total probability was considered as the most important exhibition of difference between probabilsitic
laws for classical and quantum systems. However, in papers of some authors ,e.g. Ref. 2, 5, 6, 7, 8, 13 there was pointed out that violation of the formula of total probability is just an exhibition of contextuality of quantum probabilities.

In this paper we predict that contextual statistics produced by experiments of two slit type is not reduced to
classical and quantum. Besides the absence of interference and the quantum trigonometric interference, we predict
a new type of interference -- the hyperbolic interference. In our approach it is very easy to find the type of interference
of probabilities. In a statistical test for some context $C$ we calculate the coefficient
$$
\lambda(a=x/b, C)=\frac{p_C^b(x) - p^{b/a}(x/1)p_C^a(1) - p_C^a(2) p^{b/a}(x/2) }
{2\sqrt{p^{b/a}(x/1)p_C^a(1) p_C^a(2) p^{b/a}(x/2)}}.
$$
An empirical situation with $ \lambda(a=x/b, C) > 1$ would yield evidence for quantum-like
hyperbolic behavior.  The coefficient $\lambda(a=x/b, C)$ can be easily calculated on the basis of statistical data.

Thus our hyperbolic quantum mechanics predicts a testable result, namely the hyperbolic interference,
that ordinary quantum mechanics does not!

We wrote about experiments of ``two slit type''. They need not be precisely experiments with space-variables.
The $a$ and $b$ can be any pair of incompatible observables. Incompatibility is understood as the impossibility
to escape mutual disturbances in the process of measurement. The coefficient $\lambda(a=x/b, C)$  gives the measure of
statistical disturbance. Classical measurements are characterized by (statistically) negligibly small
mutual disturbances, so here $\la(b=x/a,C)=0$ (and we have the conventional formula of total probability). Quantum measurements are characterized by mutual disturbances which are not negligible (statistically). Here
$\la(b=x/a,C)\in (0,1].$ The conventional formula of total probability is violated and we have the conventional trigonometric interference. However, the quantum case, i.e., $\la(b=x/a,C)\in (0,1],$ does not describe all nonclassical measurements. There can exist incompatible observables which produce mutual disturbances which are (statistically)
essentially larger than the conventional quantum disturbances. In such a case $\lambda(a=x/b, C) >1.$  As in the quantum case, the conventional formula of total probability is violated, but we have nonconventional hyperbolic interference.

Thus hyperbolic interference might be found in experiments with systems which are essentially more sensitive to
disturbance effects of measurement devices than quantum systems. So to find such an interference we should
go to new scales of space, time and energy: distances and time intervals which are essentially smaller than
approached in the conventional quantum experiments. One may speculate that there can be some connections  with string theory and cosmology. It may be that quantum mechanics for string theory and cosmology is  hyperbolic quantum mechanics.

Another possibility to find hyperbolic interference (which looks more realizable at the present technological level)
is to look for observables  on ordinary quantum or classical systems which would produce very strong
statistical  disturbances.

Since we derived the hyperbolic (as well as the conventional trigonometric) interference in the general contextual
probabilistic approach, our formalism can be applied to any kind of systems, for example
cognitive systems. Experiments of the two slit type can be done for cognitive systems, e.g. human beings.
Here observables $a$ and $b$ are given in the form of questions.
It might be that cognitive systems can produce hyperbolic interference and should be described
by hyperbolic quantum mechanics.

\bigskip

1. A. Yu. Khrennikov, J. Math. Phys. {\bf 44}, 2471 (2003).

2.  L. Accardi, Phys. Rep. {\bf 77}, 169(1981);
``The probabilistic roots of the quantum mechanical paradoxes,''
in {\it The wave--particle dualism:  A tribute to Louis de Broglie on his 90th
Birthday}, S. Diner, D. Fargue, G. Lochak, and F. Selleri, eds.
(D. Reidel Publ. Company, Dordrecht,  1984), pp. 47-55.

3. L. Accardi, {\it Urne e Camaleoni: Dialogo sulla realta,
le leggi del caso e la teoria quantistica} (Il Saggiatore, Rome, 1997); ``Locality
and Bell's inequality'', Q. Prob. White Noise Anal. {\bf  13},1 (2001).

4. L. Accardi, A. Fedullo, Lettere al Nuovo Cimento {\bf 34,} 161-172  (1982);
L. Accardi, Il Nuovo Cimento B {\bf 110}, 685 (1995).

5.  L. E. Ballentine,  Rev. Mod. Phys. {\bf 42},  358 (1970);
{\it Quantum mechanics} (Englewood Cliffs,
New Jersey, 1989); ``Interpretations of probability and quantum theory,''
Q. Prob. White Noise Anal. {\bf  13},  71 (2001).

6.  L. E. Ballentine, {\it Quantum mechanics} (WSP,  Singapore, 1998).

7. W. M. De Muynck, ``Interpretations of quantum mechanics,
and interpretations of violations of Bell's inequality'',
Q. Prob. White Noise Anal. {\bf  13},  95 (2001).

8. W. M. De Muynck, {\it Foundations of quantum mechanics, an empiricists approach} (Kluwer, Dordrecht,
2002).

9. S. P. Gudder, Trans. AMS {\bf 119},  428 (1965);
{\it Axiomatic quantum mechanics and generalized probability theory}
(Academic Press, New York, 1970).

10. S. P. Gudder, ``An approach to quantum probability,''
Quantum Prob. White Noise Anal. {\bf  13}, 147 (2001).

11. A. Land\'e, {\it Foundations of quantum theory} (Yale Univ. Press, 1955);
{\it New foundations of quantum mechanics} (Cambridge Univ. Press, Cambridge, 1968).

12.  G. W. Mackey, {\it Mathematical foundations of quantum mechanics}
(W. A. Benjamin INc, New York, 1963).

13. A. Yu. Khrennikov, {\it Interpretations of probability} (VSP Int. Sc. Publ.,
Utrecht, 1999).

14.  Feynman R. and Hibbs A., {\it Quantum Mechanics and Path Integrals}
(McGraw-Hill, New-York 1965).

\end{document}